\documentclass[twocolumn,prl,unsortedaddress,superscriptaddress,showpacs]{revtex4}

\usepackage{graphicx}
\usepackage{dcolumn}
\usepackage{amsmath}
\usepackage{url}
\begin{document}

\title{First Observation of the Greisen-Zatsepin-Kuzmin Suppression}

\author{R.U.~Abbasi}
\affiliation{University of Utah, Department of Physics, Salt Lake
  City, UT, USA} 

\author{T.~Abu-Zayyad}
\affiliation{University of Utah, Department of Physics, Salt Lake
  City, UT, USA} 

\author{M.~Allen}
\affiliation{University of Utah, Department of Physics, Salt Lake
  City, UT, USA} 

\author{J.F.~Amman}
\affiliation{Los Alamos National Laboratory, Los Alamos, NM, USA}

\author{G.~Archbold}
\affiliation{University of Utah, Department of Physics, Salt Lake
  City, UT, USA} 

\author{K.~Belov}
\affiliation{University of Utah, Department of Physics, Salt Lake
  City, UT, USA} 

\author{J.W.~Belz}
\affiliation{University of Utah, Department of Physics, Salt Lake
  City, UT, USA}

\author{S.Y.~Ben~Zvi}
\affiliation{Columbia University,  Department of Physics and Nevis
  Laboratory, New York, New York, USA}

\author{D.R.~Bergman}
\email[Corresponding author: ]{bergman@physics.rutgers.edu}
\affiliation{Rutgers University --- The State University of New
  Jersey, Department of Physics and Astronomy, Piscataway, NJ, USA}

\author{S.A.~Blake}
\affiliation{University of Utah, Department of Physics, Salt Lake
  City, UT, USA}

\author{O.A.~Brusova}
\affiliation{University of Utah, Department of Physics, Salt Lake
  City, UT, USA}

\author{G.W.~Burt}
\affiliation{University of Utah, Department of Physics, Salt Lake
  City, UT, USA}

\author{C.~Cannon}
\affiliation{University of Utah, Department of Physics, Salt Lake
  City, UT, USA}

\author{Z.~Cao}
\affiliation{University of Utah, Department of Physics, Salt Lake
  City, UT, USA}

\author{B.C.~Connolly}
\affiliation{Columbia University,  Department of Physics and Nevis
  Laboratory, New York, New York, USA}

\author{W.~Deng}
\affiliation{University of Utah, Department of Physics, Salt Lake
  City, UT, USA}

\author{Y.~Fedorova}
\affiliation{University of Utah, Department of Physics, Salt Lake
  City, UT, USA}

\author{C.B.~Finley}
\affiliation{Columbia University,  Department of Physics and Nevis
  Laboratory, New York, New York, USA}

\author{R.C.~Gray}
\affiliation{University of Utah, Department of Physics, Salt Lake
  City, UT, USA}

\author{W.F.~Hanlon}
\affiliation{University of Utah, Department of Physics, Salt Lake
  City, UT, USA}

\author{C.M.~Hoffman}
\affiliation{Los Alamos National Laboratory, Los Alamos, NM, USA} 

\author{M.H.~Holzscheiter}
\affiliation{Los Alamos National Laboratory, Los Alamos, NM, USA}

\author{G.~Hughes}
\affiliation{Rutgers University --- The State University of New
  Jersey, Department of Physics and Astronomy, Piscataway, NJ, USA}

\author{P.~H\"{u}ntemeyer}
\affiliation{University of Utah, Department of Physics, Salt Lake
  City, UT, USA}

\author{B.F~Jones}
\affiliation{University of Utah, Department of Physics, Salt Lake
  City, UT, USA}

\author{C.C.H.~Jui}
\affiliation{University of Utah, Department of Physics, Salt Lake
  City, UT, USA}

\author{K.~Kim}
\affiliation{University of Utah, Department of Physics, Salt Lake
  City, UT, USA}

\author{M.A.~Kirn}
\affiliation{Montana State University, Department of Physics, Bozeman,
  MT , USA}

\author{E.C.~Loh}
\affiliation{University of Utah, Department of Physics, Salt Lake
  City, UT, USA}

\author{M.M.~Maestas}
\affiliation{University of Utah, Department of Physics, Salt Lake
  City, UT, USA}

\author{N.~Manago}
\affiliation{University of Tokyo, Institute for Cosmic Ray Research,
  Kashiwa, Japan}

\author{L.J.~Marek}
\affiliation{Los Alamos National Laboratory, Los Alamos, NM, USA} 

\author{K.~Martens}
\affiliation{University of Utah, Department of Physics, Salt Lake
  City, UT, USA}

\author{J.A.J.~Matthews}
\affiliation{University of New Mexico, Department of Physics and
  Astronomy, Albuquerque, NM, USA }

\author{J.N.~Matthews}
\affiliation{University of Utah, Department of Physics, Salt Lake
  City, UT, USA}

\author{S.A.~Moore}
\affiliation{University of Utah, Department of Physics, Salt Lake
  City, UT, USA}

\author{A.~O'Neill}
\affiliation{Columbia University,  Department of Physics and Nevis
  Laboratory, New York, New York, USA}

\author{C.A.~Painter}
\affiliation{Los Alamos National Laboratory, Los Alamos, NM, USA} 

\author{L.~Perera}
\affiliation{Rutgers University --- The State University of New
  Jersey, Department of Physics and Astronomy, Piscataway, NJ, USA} 

\author{K.~Reil}
\affiliation{University of Utah, Department of Physics, Salt Lake
  City, UT, USA}

\author{R.~Riehle}
\affiliation{University of Utah, Department of Physics, Salt Lake
  City, UT, USA}

\author{M.~Roberts}
\affiliation{University of New Mexico, Department of Physics and
  Astronomy, Albuquerque, NM, USA }

\author{D.~Rodriguez}
\affiliation{University of Utah, Department of Physics, Salt Lake
  City, UT, USA}

\author{N.~Sasaki}
\affiliation{University of Tokyo, Institute for Cosmic Ray Research,
  Kashiwa, Japan}

\author{S.R.~Schnetzer} 
\affiliation{Rutgers University --- The State University of New
  Jersey, Department of Physics and Astronomy, Piscataway, NJ, USA}

\author{L.M.~Scott}
\affiliation{Rutgers University --- The State University of New
  Jersey, Department of Physics and Astronomy, Piscataway, NJ, USA}

\author{G.~Sinnis}
\affiliation{Los Alamos National Laboratory, Los Alamos, NM, USA}

\author{J.D.~Smith}
\affiliation{University of Utah, Department of Physics, Salt Lake
  City, UT, USA}

\author{P.~Sokolsky}
\affiliation{University of Utah, Department of Physics, Salt Lake
  City, UT, USA}

\author{C.~Song}
\affiliation{Columbia University, Department of Physics and Nevis
  Laboratory, New York, New York, USA}

\author{R.W.~Springer}
\affiliation{University of Utah, Department of Physics, Salt Lake
  City, UT, USA}

\author{B.T.~Stokes}
\affiliation{University of Utah, Department of Physics, Salt Lake
  City, UT, USA}

\author{S.B.~Thomas}
\affiliation{University of Utah, Department of Physics, Salt Lake
  City, UT, USA}

\author{J.R.~Thomas}
\affiliation{University of Utah, Department of Physics, Salt Lake
  City, UT, USA}

\author{G.B.~Thomson}
\affiliation{Rutgers University --- The State University of New
  Jersey, Department of Physics and Astronomy, Piscataway, NJ, USA}

\author{D.~Tupa}
\affiliation{Los Alamos National Laboratory, Los Alamos, NM, USA}

\author{S.~Westerhoff}
\affiliation{Columbia University, Department of Physics and Nevis
  Laboratory, New York, New York, USA}

\author{L.R.~Wiencke}
\affiliation{University of Utah, Department of Physics, Salt Lake
  City, UT, USA}

\author{X.~Zhang}
\affiliation{Columbia University, Department of Physics and Nevis
  Laboratory, New York, New York, USA}

\author{A.~Zech}
\affiliation{Rutgers University --- The State University of New
  Jersey, Department of Physics and Astronomy, Piscataway, NJ, USA}

\collaboration{The High Resolution Fly's Eye Collaboration}

\begin{abstract}
  The High Resolution Fly's Eye (HiRes) experiment has observed the
  Greisen-Zatsepin-Kuzmin suppression (called the GZK cutoff) with a
  statistical significance of five standard deviations.  HiRes'
  measurement of the flux of ultrahigh energy (UHE) cosmic rays shows
  a sharp suppression at an energy of $6 \times 10^{19}$ eV,
  consistent with the expected cutoff energy.  We observe the
  ``ankle'' of the cosmic-ray energy spectrum as well, at an energy of
  $4 \times 10^{18}$ eV.  We describe the experiment, data collection,
  analysis, and estimate the systematic uncertainties.  The results
  are presented and the calculation of the statistical significance of
  our observation is described.
\end{abstract}


\maketitle


In 1966, Greisen \cite{Greisen-1966-PRL-16-748}, and Zatsepin and
Kuzmin \cite{Zatsepin-1966-JETPL-4-78}, proposed an upper limit to the
cosmic-ray energy spectrum. Their predictions were based on the
assumption of a proton dominated extra-galactic cosmic-ray flux which
would interact with the photons in the cosmic microwave background
(CMB) via photo-pion production.  From the temperature of the CMB and
the mass and width of the $\Delta^+$ resonance, a ``GZK'' threshold of
$\sim{6}\times{10}^{19}$~eV was calculated, and a suppression in the
cosmic-ray flux beyond this energy (commonly called the GZK cutoff)
was predicted.  This is a strong energy-loss mechanism that limits the
range of cosmic protons above this threshold to less than $\sim50$
Mpc.

Several earlier experiments
\cite{Linsley-1963-PRL-10-146,Bird-1995-ApJ-441-144,Lawrence-1991-JPG-17-733,Pravdin-1999-ICRC-26-3-292}
have reported the detection of one event each above ${10}^{20}$~eV. A
continuing, unbroken energy spectrum beyond the predicted GZK
threshold was later reported by a larger experiment, the Akeno Giant
Air Shower Array (AGASA)
\cite{Takeda-2003-APP-19-447,Shinozaki-2007-Quarks2006}.

The High Resolution Fly's Eye (HiRes) experiment was operated on
clear, moonless nights over a period of nine years (1997-2006). During
that time, HiRes collected a cumulative exposure more than twice that
collected by AGASA above the GZK threshold.  The HiRes experiment
observes cosmic rays by imaging the extensive air shower (EAS)
generated by a primary cosmic ray.  Ultraviolet fluorescence (UV)
light is emitted by nitrogen molecules in the wake of the EAS and
collected by our detector.

Forty years after its initial prediction, the GZK cutoff has been
observed for the first time by the HiRes experiment. In this article
we describe our measurement of the flux of cosmic rays, the resulting
cosmic-ray energy spectrum, our analysis of this spectrum to infer the
existence of the cutoff, and our estimate of systematic uncertainties.
 

The HiRes project has been described previously
\cite{AbuZayyad-1999-ICRC-26-5-349,Boyer-2002-NIMA-482-457}.  The
experiment consists of two detector stations (HiRes-I and HiRes-II)
located on the U.S. Army Dugway Proving Ground in Utah, 12.6 km apart.
Each station is assembled from telescope modules (22 at HiRes-I and 42
at HiRes-II) pointing at different parts of the sky, covering nearly
$360^{\circ}$ in azimuth, and $3^{\circ}$--$17^{\circ}$ (HiRes-I), and
$3^{\circ}$--$31^{\circ}$ (Hires-II) in elevation.  Each telescope
module collects and focuses UV light from air showers using a
spherical mirror of 3.7~m$^2$ effective area. A cluster of 256
photomultiplier tubes (PMTs) is placed at the focal plane of each
mirror and serves as the camera for each telescope. The field of view
of each PMT subtends a one degree diameter cone on the sky.

HiRes data analysis is carried out in two ways. In monocular mode,
events from each detector site are selected and reconstructed
independently.  The combined monocular dataset has the best
statistical power and covers the widest energy range.  The dataset
consisting of events seen by both detectors (stereo mode data) has the
best energy resolution, but it covers a narrower energy range and has
less statistics\cite{Sokolsky-2007-ICRC-30-1262}.  This article
presents the monocular energy spectra from our two detectors.


The photometric calibration of the HiRes telescopes has been described
previously \cite{Abbasi-2005-APP-23-157}.  It is based on a portable,
high-stability ($\sim{0.5}$\%) Xenon flash lamp carried to each mirror
on a monthly basis.  Relative nightly calibrations were performed
using Yag laser light brought to each cluster of PMT's through optical
fibers.  In addition, the overall optical calibration of the HiRes
detectors is validated by reconstructing scattered light from a pulsed
laser fired into the atmosphere from locations that surround, and are
within $\sim{3.5}$~km, of the two detector sites.  We achieve
$\sim{10}$\% RMS accuracy in our photometric scale.

We monitor the UV transmission properties of the atmosphere to make a
correction for the attenuation of fluorescence light.  Steerable
lasers fire patterns of shots that cover the aperture of our
fluorescence detectors, and the detectors measure the intensity of the
scattered light.  The most important parameter we measure is the
vertical aerosol optical depth (VAOD).  The mean value of the VAOD is
0.04 with an RMS variation of 0.02.  An event at 25 km from a HiRes
detector has an average aerosol correction of $\sim$15\% upward in
energy.  Because 2.5 years of early HiRes-I data were collected before
the lasers were deployed, the spectra presented here are calculated
using a constant-atmosphere assumption, using the measured average
value for the VAOD.  We have tested this assumption by calculating the
energy spectrum from our later data, using the actual hourly
measurements.  Comparing the resulting spectra from the two methods,
we obtain flux values that agree to within a few percent
\cite{Abbasi-2007-APP-astroph-0607094}.

Another important parameter in our analysis is the fluorescence yield
(FY): the number of photons generated per ionizing particle per unit
path length.  FY measurements have been made by several groups
\cite{Bunner-1967-Thesis,Kakimoto-1996-NIMA-372-527,Nagano-2003-APP-20-293,Belz-2006-APP-25-129}.
For the energy spectrum determination used in this paper, we have used
the spectral shape of Bunner \cite{Bunner-1967-Thesis} and the
integral yield reported by Kakimoto {\it et al.}
\cite{Kakimoto-1996-NIMA-372-527}. Our systematic studies have shown
that this set of assumptions produces absolute fluorescence flux
values that are equal, within $\sim6\%$, of those obtained using a fit
to all the results cited \cite{2007-FY-measurements}.


The details of HiRes event selection have been described previously
\cite{Abbasi-2004-PRL-92-151101,Abbasi-2005-PLB-619-271}.  An
additional cut on the distance to showers has been applied in the
HiRes-II data collected after those shown in
\cite{Abbasi-2005-PLB-619-271}.  This cut is applied to make the
aperture (defined as the product of collection area and solid angle)
calculation more robust.  The event reconstruction procedure begins
with the determination of the shower axis.  A plane containing the
axis of the shower and the detector, the shower-detector plane (SDP)
is determined from the pointing direction of triggered PMTs.  For the
HiRes-II monocular dataset, the PMT times are then used to find the
distance to the shower and the angle, $\psi$, of the shower within the
SDP.  This timing fit measures $\psi$ to an accuracy of
$\sim5^{\circ}$ RMS.
 
The number of shower particles as a function of atmospheric depth is
then determined.  This calculation uses the FY and corrects for
atmospheric attenuation.  We fit this shower profile to the
Gaisser-Hillas function \cite{Gaisser-Hillas-1977-ICRC-15-8-353},
after having subtracted scattered \v{C}erenkov light produced by the
air shower particles.  This profile fit yields both the energy of the
shower and the depth at the shower maximum, $X_{\rm max}$.  A typical
HiRes profile is displayed in \cite{Abbasi-2005-APP-23-157}.  The
energy resolution of the HiRes-II detector is about 12\% at high
energies.

\begin{figure}[ht]
  \begin{center}
    \includegraphics[width=\columnwidth]{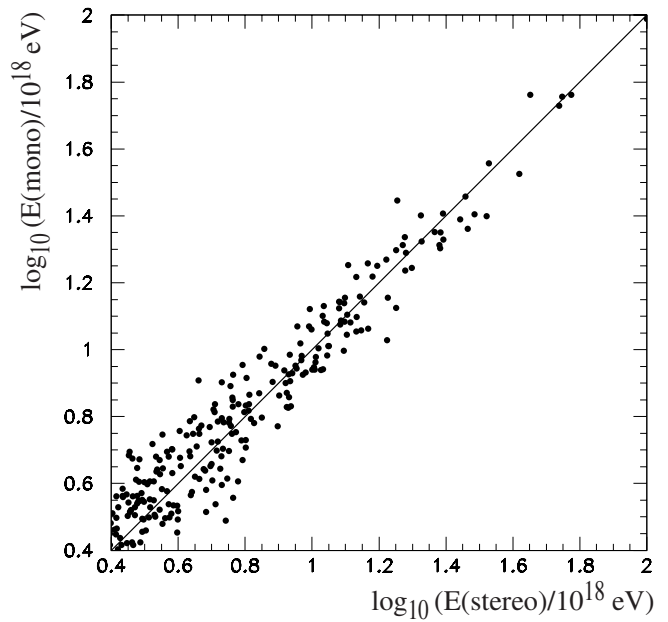}
  \end{center}
  \vspace{-0.25in}
  \caption{HiRes-I energies calculated with the event geometry
    reconstructed in monocular mode using the profile-constrained fit
    vs. the energy reconstructed in stereo mode.}
  \label{fig:mono}
\end{figure}

The HiRes-I detector, with its limited elevation coverage, does not
typically observe enough of the shower for a reliable timing fit.  For
this reason the HiRes-I monocular reconstruction combines the timing
and profile fits in a profile-constrained fit (PCF).  The PCF
reconstructs $\psi$ with an accuracy of $\sim7^{\circ}$ RMS.  The PCF
has been validated by comparing the PCF energies to those found using
stereo geometries in that subset of the data observed by both
detectors as shown in Figure~\ref{fig:mono}.  The energy resolution of
the HiRes-I detector is about 17\% at high energies.

Finally, a correction is made for the energy carried by shower
components which do not deposit their energy in the atmosphere.  This
correction includes primarily the energy of neutrinos and muons that
strike the earth.  The correction is calculated using shower
simulations in CORSIKA \cite{Heck-1998-FZKA-6019} with hadronic
interaction simulated by QGSJet \cite{Kalmykov-1997-NPBps-52b-17}.
The correction is $\sim10\%$.  Simulations using Sibyll
\cite{Fletcher-1994-PRD-50-5719} find a correction within 2\%
\cite{Abbasi-2007-APP-astroph-0607094} of that found via QGSJet.


The measurement of the cosmic-ray flux requires a reliable
determination of the detector aperture.  The aperture of the HiRes
detectors has been calculated using a full Monte Carlo (MC)
simulation.  The MC includes simulation of shower development (using
CORSIKA), fluorescence and \v{C}erenkov light production, transmission
of light through the atmosphere to the detector, collection of light
by the mirrors, and the response of the PMTs, electronics and trigger
systems.  Simulated events are recorded in the same format as real
data and processed in an identical fashion.  To minimize biases from
resolution effects, MC event sets are generated using the published
measurements of the energy spectrum \cite{Bird-1993-PRL-71-3401} and
composition
\cite{AbuZayyad-2000-PRL-19-4276,AbuZayyad-2001-ApJ-557-686,Abbasi-2005-ApJ-622-910}.

\begin{figure}[tbh]
  \begin{center}
    \includegraphics[width=\columnwidth]{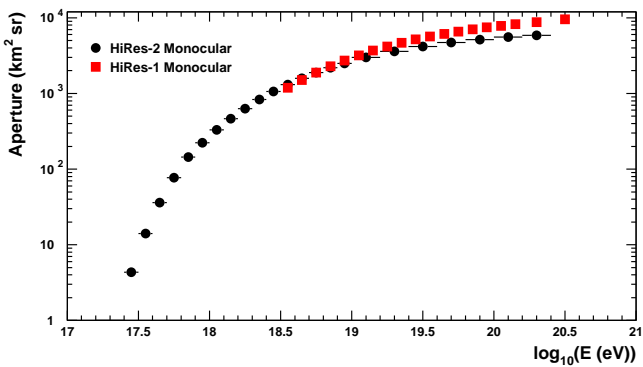}
  \end{center}
  \vspace{-0.25in}
  \caption{The apertures (defined as the product of collection area
    and solid angle) of the HiRes-I and HiRes-II detectors
    operating in monocular mode.}
  \label{fig:aperture}
\end{figure}

To ensure the reliability of the aperture calculation, the MC
simulation is validated by comparing key distributions from the
analysis of MC events to those from the actual data.  Several of these
comparisons were shown in reference \cite{Bergman-2007-NPBps-165}. Two
comparisons are especially noteworthy.  The data-MC comparison of the
distances to showers shows that the simulation accurately models the
coverage of the detector.  The comparison of event brightness shows
that the simulations of the optical characteristics of the detector,
and of the trigger and atmospheric conditions, accurately reproduce
the data collection environment.  The excellent agreement between the
observed and simulated distributions shown in these cases is typical
of MC-data comparisons of other kinematic and physical quantities, and
this agreement demonstrates that we have a reliable MC simulation
program and aperture calculation.  Figure \ref{fig:aperture} shows the
result of the aperture calculation for both HiRes-I and HiRes-II in
monocular mode.

\begin{figure}[tbh]
  \begin{center}
    \includegraphics[width=\columnwidth]{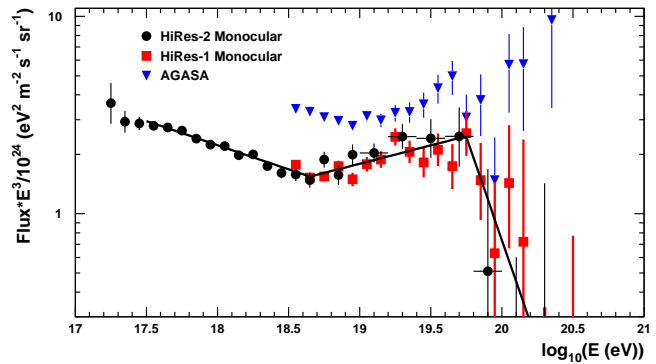}
  \end{center}
  \vspace{-0.25in}
  \caption{The cosmic-ray energy spectrum measured by the HiRes
    detectors operating in monocular mode.  The spectrum of the
    HiRes-I and HiRes-II detectors are shown.  The highest two energy
    bins for each detector are empty, with the 68\% confidence level
    bounds shown.  The spectrum of the AGASA experiment is also
    shown \cite{Takeda-2003-APP-19-447,Shinozaki-2007-Quarks2006}.}
  \label{fig:spectrum}
\end{figure}

Figure \ref{fig:spectrum} shows the monocular energy spectra from the
two HiRes detectors \cite{Bergman-2007-WebSpectra}.  The data included
in the figure were collected by HiRes-I from May, 1997 to June, 2005,
and by HiRes-II from December, 1999 to August, 2004.  Figure
\ref{fig:spectrum} shows the flux multiplied by $E^3$, which does not
change the statistical interpretation of the results but highlights
features more clearly.  Two prominent features seen in the figure are
a softening of the spectrum at the expected energy of the GZK
threshold of $10^{19.8}$~eV, and the dip at $10^{18.6}$ eV, commonly
known as the ``ankle''.  Theoretical fits to the spectrum
\cite{Berezinsky-2006-PRD-74-043005} show that the ankle is likely
caused by $e^+e^-$ pair production in the same interactions between
CMB photons and cosmic-ray protons where pion production produces the
GZK cutoff.  The observation of both features is consistent with the
published HiRes results of a predominantly light composition above
$10^{18}$~eV \cite{Abbasi-2005-ApJ-622-910}.

At lower energies, the cosmic-ray spectrum is well fit by a piece-wise
power law model. A similar fit also gives an excellent representation
of the spectrum in Figure~\ref{fig:spectrum}. The three straight line
segments shown represent the result of a fit of the measured flux to a
triple-power law. The fit contains six free parameters: one
normalization, the energies of two floating break points, and three
power law indices.

We performed a binned maximum likelihood fit
\cite{Yao-2006-JPG-33-302} to the data from the two detectors.  The
fits include two empty bins for each monocular dataset. We found the
two breaks at $\log E$ ($E$ in eV) of $19.75 \pm 0.04$ and $18.65 \pm
0.05$, corresponding to the GZK cutoff and the ankle, respectively.
When the datasets were made statistically independent by removing
events seen by both detectors from the HiRes-I dataset, we obtained a
$\chi^2$ of 35.1 in this fit for 35 degrees of freedom (DOF).  In
contrast, a fit to a model with only one break point, while able to
locate the ankle (at the same energy), yielded a $\chi^2$/DOF=63.0/37
\cite{2007-Auger-Break}.

A measure of the significance of the break in the spectral index at
$10^{19.8}$ eV can be made by comparing the actual number of events
observed above the break to the expected number for an unbroken
spectrum. For the latter, we assume the power law of the middle
segment to continue beyond the threshold. From the independant HiRes
exposures (with events seen by both detectors removed from HiRes-I),
we expect 43.2 events above $10^{19.8}$ eV from the extrapolation,
whereas 13 events were actually found in the data.  The Poisson
probability for the observed deficit is $7\times10^{-8}$, which
corresponds to 5.3 standard deviations.  We conclude that we have
observed the GZK cutoff with a 5 standard deviation significance.


One question that remains is whether the sources of extragalactic UHE
cosmic rays have properties that could change the GZK energy.  A study
by V. Berezinsky {\it et al.}  \cite{Berezinsky-2006-PRD-74-043005}
found that the density of sources in the local area should change the
power law of the energy spectrum above the GZK cutoff, but not the GZK
energy itself.  The average power law of the sources could change the
GZK energy somewhat, but the $E_{1/2}$ method suggested by V.
Berezinsky and S. Grigorieva \cite{Berezinsky-Grigoreva-1988-AA-199-1}
provides a test of whether a break is the GZK cutoff independent of
power law over a wide range.  $E_{1/2}$ refers to the energy at which
the integral energy spectrum falls to half of what would be expected
in the absence of the GZK cutoff.  To calculate $E_{1/2}$ we used the
HiRes monocular energy spectra and the integral of the power law
spectrum used above to estimate the number of expected events above
the break.  We find $E_{1/2} = 10^{19.73\pm0.07}$.  Berezinsky and
Grigorieva predict a robust theoretical value for $E_{1/2}$ of
$10^{19.76}$ eV for a wide range of spectral slopes
\cite{Berezinsky-Grigoreva-1988-AA-199-1}. These two values are
clearly in excellent agreement, supporting our interpretation of the
break as the GZK cutoff.

We measure the index of the power law to be $3.25\pm0.01$ below the
ankle, $2.81\pm0.03$ between the ankle and the GZK cutoff, and
$5.1\pm0.7$ above the GZK cutoff.


For the monocular analyses, the main contributions to the systematic
uncertainty in the energy scale and flux measurements are: PMT
calibration (10\%), fluorescence yield (6\%), missing energy
correction (5\%), aerosol component of the atmospheric attenuation
correction (5\%), and mean energy loss rate estimate (the flux of
fluorescence photons is proportional to the mean $dE/dx$ of the
particles in the shower \cite{Belz-2006-APP-25-57}) (10\%).  Since
these uncertainties arise from very different sources, we add them in
quadrature, giving a total energy scale uncertainty of 17\%, and a
systematic uncertainty in the flux of 30\%.


In summary, we have measured the flux of ultrahigh energy cosmic rays
with the fluorescence technique, in the energy range $10^{17.2}$ to
above $10^{20.5}$ eV.  We observe two breaks in the energy spectrum
consistent with the GZK cutoff and the ankle.  The statistical
significance of the break identified with the GZK cutoff is 5 standard
deviations.  We measure the energy of the GZK cutoff to be
$(5.6\pm0.5\pm0.9)\times10^{19}$~eV, where the first uncertainty is
statistical and the second is systematic.

This work is supported by US NSF grants PHY-9321949, PHY-9322298,
PHY-9904048, PHY-9974537, PHY-0098826, PHY-0140688, PHY-0245428,
PHY-0305516, PHY-0307098, and by the DOE grant FG03-92ER40732.We
gratefully acknowledge the contributions from the technical staffs of
our home institutions. The cooperation of Colonels E.~Fischer,
G.~Harter and G.~Olsen, the US Army, and the Dugway Proving Ground
staff is greatly appreciated.

\bibliography{UHECR-spectrum}

\begin{thebibliography}{35}
\bibitem[{200({\natexlab{a}})}]{2007-FY-measurements}
\bibinfo{note}{There are several new measurements of the FY about to be
  published which may change the average FY.}
\end{thebibliography}

\end{document}